\def\year{2015}
\RequirePackage[final]{graphicx} 
\documentclass[letterpaper,draft]{article}
\usepackage{lipsum}
\usepackage{url}
\usepackage{amsmath, amsfonts, amssymb, amsthm, mathrsfs}
\usepackage{tikz}
\usetikzlibrary{shapes,snakes}
\usetikzlibrary{positioning,decorations.pathreplacing,shapes,arrows,automata}
\usepackage{caption} 
\usepackage{filecontents}
\usepackage{wrapfig}
\usepackage{multicol}
\usepackage{pgfplots}
\pgfplotsset{compat=newest}
\usepgfplotslibrary{units}
\usepackage{xcolor}
\usepackage{subfigure}
\usepackage{array}
\newcolumntype{C}[1]{>{\centering\arraybackslash}p{#1}}
\usetikzlibrary{arrows,automata,trees}
\usepackage{adjustbox}
\usepackage{multirow}
\usepackage{colortbl}
\definecolor{kugray5}{RGB}{224,224,224}
\usepackage{rotating}
\newcolumntype{a}{>{\columncolor{green}}c}
\newcolumntype{o}{>{\columncolor{orange}}c}
\definecolor{mycolorG}{rgb}{.4,1,.4}
\definecolor{mycolorB}{rgb}{.59,1,1}
\definecolor{mycolorR}{rgb}{1,.38,.01} 
\usetikzlibrary{decorations.pathreplacing}

\usepackage{sidecap}
\usepackage{aaai}
\usepackage{times}
\usepackage{helvet}
\usepackage{courier}
\graphicspath{{fig/}}
\usepackage{ifdraft}                                               
\usepackage[normalem]{ulem}                                        
\ifdraft{ 
\usepackage[textwidth=1.5cm,textsize=footnotesize,color=white,bordercolor=red,linecolor=none,colorinlistoftodos]{todonotes}
\newcommand{\mh}[1]{{\scriptsize\color{red}[MH: #1]}}
}{ 
\newcommand{\mh}[1]{}

}
\usepackage[inline]{enumitem}

\newcommand{\fref}[1]{Fig.~\ref{#1}}       
\newcommand{\sref}[1]{\S\ref{#1}}          
\newcommand{\tref}[1]{\tablename~\ref{#1}} 
\newcommand{\eref}[1]{(\ref{#1})}          

\newcommand{\ie}[1]{\textit{i.e.,}~#1} %

\newcommand{\x}       {s}             
\newcommand  {\y}     {o}                      
\renewcommand{\u}     {a}             
\renewcommand{\j}     {r}             
\newcommand{\xp}      {\x'}           

\newcommand  {\xd}    {\x^*}                   

\newcommand{\X}       {\MakeUppercase{\x}}     
\newcommand{\Y}       {\mathcal{\MakeUppercase{\y}}} 
\newcommand{\U}       {\MakeUppercase{\u}}     
\newcommand{\F}       {T}                      
\newcommand{\J}       {R}                      

\newcommand{\Uh}      {\mathscr{\MakeUppercase{\u}}}           
\newcommand{\Um}      {\mathbb{\MakeUppercase{\u}}}  
\newcommand{\Un}      {\mathcal{\MakeUppercase{\u}}}           

\newcommand{\mdp}       {MDP}       
\newcommand{\mdps}       {MDPs}       
\newcommand{\V}       {V}       
\newcommand{\po}       {PO}       
\newcommand{\Om}       {\Omega}       
\newcommand{\Bss}       {B}       
\newcommand{\bs}       {b}       

\newcommand{\Tw}          {T}              
\newcommand{\nP}            {R}                
\newcommand{\inv}         {I}               
\newcommand{\mt}         {t}                
\newcommand{\ac}         {a}               
\newcommand{\cc}         {c}                

\newcommand{\G}          {G}                      

\newcommand{\zt}         {\zeta}             

\newcommand{\ca}         {\alpha}                 
\newcommand{\cb}         {\beta}                  

\newcommand{\dn}        {\circ}              
\newcommand{\uba}        {\uparrow}              
\newcommand{\ubb}        {\rightarrow}            
\newcommand{\usa}        {\downarrow}            
\newcommand{\usb}        {\leftarrow}             

\newcommand{\usam}        {\Uparrow}           
\newcommand{\usbm}        {\Rightarrow}        
\newcommand{\ubam}        {\Downarrow}       
\newcommand{\ubbm}        {\Leftarrow}           

\begin{document}

\title{Learning Unfair Trading: a Market Manipulation Analysis From the Reinforcement
Learning Perspective}
\author{Enrique~Mart\'{i}nez-Miranda \and Peter McBurney \and Matthew~J.~Howard \\
Department of Informatics\\
King's College London\\
\{enrique.martinez\_miranda,peter.mcburney,matthew.j.howard\}@kcl.ac.uk}
\maketitle

\begin{abstract}
\begin{quote}
Market manipulation is a strategy used by traders to alter the price of financial
securities. One type of manipulation is based on the process of buying or selling assets
by using several trading strategies, among them \textit{spoofing} is a popular strategy 
and is considered illegal by market regulators. Some promising tools have been 
developed to detect manipulation, but cases can still be found in the markets. In this 
paper we model \textit{spoofing} and \textit{pinging} trading, two strategies that differ in 
the legal background but share the same elemental concept of market manipulation. 
We use a reinforcement learning framework within the full and partial observability of 
Markov decision processes and analyse the underlying behaviour of the manipulators 
by finding the causes of what encourages the traders to perform fraudulent activities. 
This reveals procedures to counter the problem that may be helpful to market regulators 
as our model predicts the activity of spoofers.
\end{quote}
\end{abstract}

\section{\label{s:intro}Introduction}

Market microstructure is a branch of finance concerned with analysis of the trading 
process arising from the exchange of assets under a given set of rules~\cite{ohara98}. 
In double auction markets, this exchange of assets is done when the buy and sell sides 
agree on the amount to pay/receive for the trade, but this agreement depends on the 
different strategies implemented by both sides. A trading strategy is by itself a plan of 
actions designed to achieve profitable returns by buying or selling financial
assets~\cite{pardo08}.

While trading strategies are meant to follow the well established rules of the markets, 
some traders prefer to misbehave and take advantage of others by manipulating the
price of the assets being traded. For instance, some traders can manipulate by 
spreading false information to other market participants or by taking actions that may 
affect the perceived price~\cite{allenG92}, just as the case of the strategy called
\textit{pump and dump}. Others, on the contrary, prefer to take actions 
directly involved in the exchange of the assets by artificially inflating or deflating the 
price in order to obtain profits. Several manipulative strategies based on the trading 
process are well known in the financial \textit{argot} and have names like 
\textit{ramping}, \textit{wash trading}, \textit{quote stuffing}, \textit{layering}, 
\textit{spoofing}, among others. \textit{Spoofing} is one of the most popular strategies 
that uses \textit{non-bona fide} orders to improve the price and is considered illegal by 
market regulators~\cite{aktas13}. A similar strategy used by high-frequency traders 
(HFTs) is called \textit{pinging} where HFTs place orders without the intention of 
execution, but to find liquidity not displayed in the \textit{order book} (where all buy and 
sell orders are listed in double auction markets), and has caused controversy as it can 
be viewed as a manipulative strategy~\cite{scopino15}.

Studies found in the literature that analyse the problem of market manipulation have
mainly focused on the development of methods for detection. However, there has been 
little analysis on the behaviour of market manipulators, an area that may reveal the 
cause of why these economic agents take such actions, thus examining this might be 
helpful for market regulators to develop counter-measures that may discourage or 
preclude fraudulent strategies

We propose to model spoofing and pinging strategies in the context of \textit{portfolio} 
\textit{growth} maximisation, \ie{the expected capital appreciation over time 
of an investment account}. We use a reinforcement learning agent that simulates the 
behaviour of the \textit{spoofing trader} in the context of Markov decision processes 
(\mdp), while a partially observable \mdp\ is used to model the \textit{pinging trader} 
since the latter involves hidden state in the order book. We use a fixed environment 
where transitions and rewards do not change in time, but the agent has the option to 
transition between ``two different'' state representations that, both combined, are the 
full state representation of the environment that simulates the manipulation process.

Our contribution is to show how  these manipulative trading strategies can be modelled 
in a (\po)\mdp\ framework and how this reveals the causes of market manipulation in 
terms of the incentives present in the market, and the dynamics of how it operates. 
From this, we aim to examine two main questions: i) can spoofing and pinging modelled
by and \mdp\ and \po\mdp\ respectively, be optimal strategies when compared to 
\textit{honest} behavior while seeking for growth maximisation? ii) If the manipulative 
strategies are optimal, which mechanisms can market regulators implement in order to 
discourage or disincent traders taking such behaviour? The results of this yield 
recommendations to market regulators as to how to stop manipulative behaviour.

\section{\label{sec:literatureReview}Related Work}

Research on price manipulation has been done using several approaches. Some 
authors have developed analytical models with the intention to investigate manipulative 
strategies performed by \textit{large traders} under the hypothesis of stochastic 
economies with finite/infinite horizon and time dependent price processes
\cite{jarrow92MM}. Others take a continuous-time economy with risky and risk free 
assets and different agents involved in a game where \textit{predatory trading} (trading 
style that takes advantage of other investors' needs) leads to price overshooting and 
amplifies the selling cost and default risk of large traders~\cite{brunnermeier05}. Others 
consider the problem where manipulative uninformed traders can profit by selling a 
given firm's stock, thus providing a starting point to restrict \textit{short selling} (when 
traders sell a security not owned)~\cite{goldsteinG08}.

Other researchers have focused in the application of data driven approaches with the 
aim to present empirical evidence of stock price manipulation under the assumption of 
the presence of arbitrageurs or information seekers acting rationally
\cite{aggarwalW06} or by finding unusual patterns of trading activities and systematic 
profitability based on \textit{market timing} and \textit{liquidity} performed by brokers in 
emerging markets~\cite{khwajaM05}. An agency-based model is tested
with empirical data where brokers manipulate the closing price to influence his 
customer's perception about his performance~\cite{hillionS04}.

Also, behavioural stances have been mixed with  theoretical and data driven
approaches. An analytical framework is developed that describes trade-based 
manipulation as an intentional act to produce changes in the price and obtain a profit, 
so one could clarify what does and does not constitute manipulation
\cite{ledgerwoodC2012}. Evidence of trade-based manipulation and its effects on 
investor behaviour and market efficiency is provided, where the manipulator pretends to 
act as an informative trader that may affect the reaction of other investors
\cite{kongW14}.

Furthermore,  discriminative models are intended to detect market manipulation based 
on empirical data. By using economic and statistical analysis it is possible to detect
manipulation \textit{ex post}, suggesting that the existence of regulatory framework may 
be inefficient~\cite{pirrong04}. Machine learning techniques have also been applied for 
detection of manipulation.  Based on trading data, some authors suggest that 
Artificial Neural Networks and Support Vector Machines are effective techniques to 
detect manipulation~\cite{hulisiDA09}. Others suggest that a method called ``hidden 
Markov model with abnormal states'' is capable to model and detect price manipulation 
patterns, but further calibration is necessary~\cite{caoLCBM13}. Data mining methods 
for detecting intraday price manipulation have been used to classify and identify 
patterns linked to market manipulation at different time scales, but further research is 
needed to address the challenge on detecting the different forms of manipulation
\cite{diazSTS11}. Furthermore, Na\"ive Bayes is a good classifier for predicting potential 
trades associated  to market manipulation~\cite{golZD14}. For the case of 
\textit{spoofing} trading, detection can be done with the implementation of supervised 
learning algorithms~\cite{caoLCBM14}, or can be identified by modelling trading 
decisions as \mdps\ and using Apprenticeship Learning to learn the reward function
\cite{yang2012behavior}.

Though research is extensive in the area of market manipulation, few develop 
generative models of what encourages these economic agents to follow the disruptive 
strategies. Furthermore, few of them provide recommendations to regulatory entities 
and/or firms~\cite{rossiDeisRP15} to encourage traders to stop this harmful behaviour. 
Different to the discriminative models that are intended to distinguish the manipulative
behaviour from other strategies, we use the (\po)\mdp\ approach to model
spoofing/pinging as it predicts the behavior of manipulators in terms of the market
conditions, thus providing a powerful tool that can be used by market regulators to 
counter the manipulative strategies.

\section{\label{s:probForm}Problem Formulation}

\subsection{\label{ss:marketManipModel}{Trading in a Bull Market}}

In this work we are focused on modelling two trade-based market manipulation 
strategies as follows. Suppose there is a trader managing an investment portfolio
in behalf of a brokerage firm and has the objective to get high trading profits that may 
produce portfolio growth in the short/medium term. Suppose the agent is trading in a
futures market and the portfolio consists of two different contracts, $\ca$ and
$\cb$, with a market full of optimism so prices are rising (a situation known as a 
\textit{bull market}). Mathematically, the capital of the investment account at given 
market \textit{tick} $\mt\in[0,\Tw]$ (where a \textit{tick} represents the execution of a 
new trade in the market, either from the trader or any other participant) can be written 
as
\begin{equation}\label{eq:investmentCap}
\inv_\mt = \ac_\mt + \cc_\mt,
\end{equation}
where $\ac_{\mt}=\ac^{\ca}_{\mt}+\ac^{\cb}_{\mt}$ is the capital associated to the
\textit{market value} of the contracts $\ca$ and $\cb$, and $\cc_{\mt}$ is the cash to be 
used for future purchases of more contracts. The variable $\ac_\mt$ changes at every 
\textit{tick} since the prices of the contracts are following a trend, while $\cc_\mt$
changes due to cash inflows/outflows (by the sale/purchase of contracts). The net 
profit of the investment over a \textit{tick window} $[0,\Tw]$ is
\begin{equation}\label{eq:netProfit}
\nP=\G_\Tw-\sum_{\mt=0}^{\Tw} \zt_\mt,
\end{equation}
where $\G_\Tw=\inv_\Tw-\inv_0$ is the investment growth, and $\zt_\mt$ are the 
direct transaction costs associated to the trading of the contracts (such as exchange 
and government fees).

Under bull market conditions, one way in which the trader can profit from the
portfolio's growth is with a simple \textit{buy and hold} strategy, an almost risk-free 
strategy whereby she purchases contracts $\ca$ and $\cb$ and simply waits, in the long 
term, for the prices to rise before selling for a profit. However, the trader may, 
alternatively, be aiming for a higher target growth $\G_{\Tw}^{*}$ in the short/medium 
term, requiring a more active strategy than the ``buy and hold'', \ie{buying and selling 
contracts $\ca$ and $\cb$, subject to the transaction costs $\zt_\mt$}.

For this, the trader can behave in several different ways. First, the trader may trade 
\textit{honestly}, \ie{following all the market rules}, by buying more contracts or selling 
them when she believes is profitable. In this way, the invested capital may appreciate 
and produce growth if such profits are larger than the direct cost associated to the 
trading process $\zt_\mt$. Alternatively, she may act as a \textit{manipulative} trader to 
control the price of the contracts in order to accelerate the growth process and quickly 
reach the desired  $\G_{\Tw}^{*}$. In either case, following the transaction, the trader 
ends up with a different proportion of the contracts $\ca$ and $\cb$, rebalancing the 
quantities $\ac_\mt$ and $\cc_\mt$ and thereby finds herself in a new level of growth
$\G_\mt$ at a given \textit{tick} $\mt$.

\begin{figure}[htp!]
\vspace{-7pt}
\centering
    \begin{tikzpicture}
    \begin{axis} [
       width=0.45\textwidth,
       height=0.2\textheight,
       legend style={at={(0.5,-0.25)}, anchor=north,legend columns=3},,
       legend style={fill=none, draw=none},
       grid=both,
       xlabel=$ticks$,
       ylabel=$Profit$
     ]
    \addplot [dotted] table [x index=0, y index=1,col sep=comma] {simulation.dat};
    \addplot [dashed] table [x index=0, y index=2,col sep=comma] {simulation.dat};
    \addplot [solid] table [x index=0, y index=3,col sep=comma] {simulation.dat};
    \addplot [only marks,mark=triangle*] coordinates { (23,10037689.34) };
    \addplot [only marks,mark=triangle*] coordinates { (46,10224375.45) };
    \addplot [only marks,mark=triangle*] coordinates { (69,10393942.3) };
    \addplot [only marks,mark=triangle*] coordinates { (92,10638931.16) };
    \addplot [only marks,mark=triangle*,mark options={fill=white}]
    coordinates { (46,10326497.54) };
    \addplot [only marks,mark=triangle*,mark options={fill=white}]
    coordinates { (92,10746842.58) };
    \legend{$Buy\&Hold$,$Honest$,$Manipulation$}
    \end{axis}
    \end{tikzpicture}
\vspace{-20pt}
\caption{A simulation of profits gained from different trading strategies during a bull 
market period.}
\vspace{-1pt}
\label{f:growthMaxim}
\end{figure}
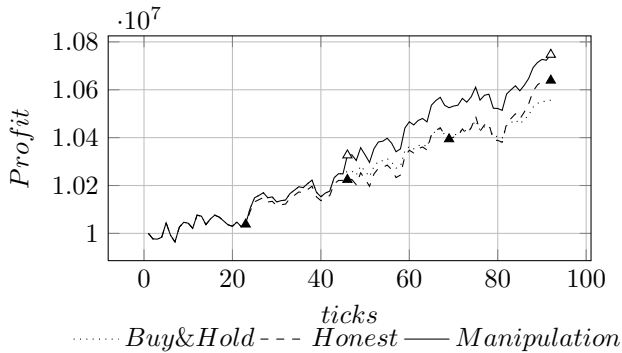

This process is illustrated in \fref{f:growthMaxim}  where the three strategies are 
simulated on closing prices determined by the market indexes \textit{S\&P 500} and 
\textit{NASDAQ Composite} for contracts $\ca$ and $\cb$, respectively, in the period 
February 27, 1995 to May 5, 1995. Initially, the trader has 1,000 contracts in both assets 
and the account's capital value is 10 million monetary units. While the market evolves, 
the trader takes actions represented by the filled (honest)/non-filled (manipulation) 
triangles, reaching new levels of profit determined by changes in the growth $\G_\mt$ 
and the payment of transaction costs, $\zt_\mt$. In our simulation, the manipulative 
strategy has the best performance, giving a signal that price manipulation is more 
effective while maximising investment growth.

However, transitioning in the different levels of growth $\G_\mt$ by trading is 
possible only when a trade is executed. In double auction markets this process can be 
performed when the buying and selling prices match, so the exchange of assets can 
proceed. This is known as \textit{liquidity} and depends on the degree of trading activity 
implemented by other market participants. In our model, honest actions taken by the 
trader can lead to no change in the level of growth $\G_\mt$ if liquidity is poor in the 
contract to trade, but a manipulator can take advantage of this situation by placing a 
large order that may gain the interest of other market participants and start a process of 
price improvement.

\begin{wrapfigure}{r}{0.12\textwidth}
\centering
\vspace{-5pt}
\includegraphics[scale=.7]{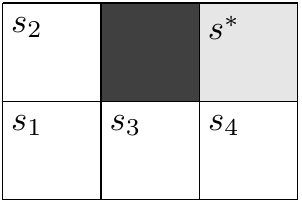}
\caption{Idealised representation illustrating the different levels of $\x_\mt$ while
maximising investment growth.}
\vspace{-5pt}
\label{f:simpleGW}
\end{wrapfigure}

An illustrative example of growth maximisation is provided in \fref{f:simpleGW}, 
where we changed the notation of $\G_\mt$ to $\x_\mt$, $\mt\in[0,4]$, and
$\G_{\Tw}^{*}$ to $\x^{*}$. There, the four growth levels correspond to holding a 
portfolio containing different proportions of contracts, for example, in $\x_1$ the 
trader holds one contract of type $\ca$ and one contract $\cb$. If the trader choses to 
buy a second $\ca$ contract, \ie{action ``Buy $\ca$'' ($\uba$)}, she transitions to 
growth  level $\x_2$ -- holding two $\ca$ contracts and one $\cb$ contract by paying the 
associated transaction costs $\zt_1$. Similarly, if she then chooses to sell the second
$\ca$ contract, \ie{action ``Sell $\ca$'' ($\usa$)}, she will return to growth level $\x_1$, 
now paying $\zt_2$ costs. These actions define honest actions for the $\ca$ contracts, 
with homologue actions for contract $\cb$ (``Buy $\cb$'' ($\ubb$) and ``Sell $\cb$''
($\usb$)).
\setlength{\columnsep}{11pt}%
Additionally, while in $\x_2$ taking actions ``Buy $\ca$'' ($\uba$) and ``Sell
$\cb$'' ($\usb$), result in no change in the level of growth. This is due to orders placed 
by the trader that were never filled because the price was too high/low while trying to 
sell/buy the $\cb$/$\ca$ contract, a process that happens in all of the edges of the grid. 
We are assuming the trader only places limit orders, \ie{orders with a fixed volume and 
price listed in the order book according to the market rules}, so the agent's orders will 
be filled only when a counterpart exists, if not, then the order is not executed and no 
transaction costs $\zt_\mt$ are added. Similarly, for action ``Buy $\cb$'' ($\ubb$) the
trader faces the problem of poor liquidity in the asset. We associate this obstacle (poor 
liquidity) for the trader to the black square in the representation of \fref{f:simpleGW}, 
with the option for the trader to try to manipulate the asset's price as a way to 
incentivize liquidity.

\subsection{\label{ss:spoofingStrategy}Price Manipulation by Spoofing}

Spoofing is an illegal trading strategy used by traders intended to manipulate the price 
of a given asset by placing large orders (spoofing orders) without the intention of
execution, but to give misleading information to other market participants in terms of the 
asset's supply and demand, thus producing a change in the price
\cite{lee2013microstructure}. Once the price is affected, the trader cancels the spoofing 
order and places the real order on the opposite trading side.

\begin{figure}[t!]
\centering
\includegraphics[scale=.8]{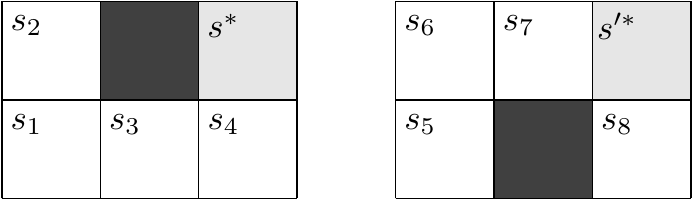}
\caption{Representation illustrating the effect of the spoofing in the process of
investment growth maximisation.}
\vspace{-10pt}
\label{f:gwSpoofing}
\end{figure}

In our model, spoofing is illustrated as follows. Consider the case that, by taking 
spoofing actions, the trader can overturn the lack of liquidity in the asset $\cb$ 
and take advantage of improved prices. In \fref{f:gwSpoofing}, this corresponds to the
obstacle switching from the top centre bin to the bottom centre bin, showing the effect of 
manipulation while purchasing more contracts (the obstacle could switch to other cells, 
but it will not allow to analyse the effects of manipulation in terms of solving an 
optimisation problem as explained in the next sections). This can yield gains for the 
trader, for example, starting in $\x_2$, the trader can take action ``Manipulative Buy
$\cb$ ($\usbm$)'' -- \ie{use spoofing to buy $\cb$, by placing a large spoofing sell order 
for $\cb$, cancelling it, and then buying $\cb$ at an improved price}. Once the obstacle 
to switch to the bottom bin, the agent finds herself in $\x_7$, closer to $\x'^*$.

The two representations in \fref{f:gwSpoofing} have the same levels of growth but with
different conditions associated to market liquidity, thus giving an idea of the effects of 
price manipulation. This effect is related to \textit{market impact}, where most of honest 
traders will avoid it as it represents indirect extra costs, but for a manipulator like a 
spoofer it represents profits that may accelerate the portfolio's growth.

\subsection{\label{ss:mdpModel}Spoofing as a Markov Decision Process}

A natural model of the scenario described in \sref{ss:spoofingStrategy}, is that of a 
Markov Decision Process (\mdp)
\cite{nevmyvaka2006reinforcement,yang2012behavior}. In general, an \mdp\ is defined 
by the tuple $\{\X,\U,\F,\J\}$, where $\X$ and $\U$ are sets of states and actions, 
respectively ($\x\in\X$ and $\u\in\U$), $\J$ is the set of rewards ($\j\in\J$), and $\F$ is a 
set of transition probabilities (\{$P(\xp|\x,\u)\}\in\F$ where $P(\xp|\x,\u)$ represents the 
probability of transitioning to state $\xp$ from $\x$ after action $\u$). Actions are taken 
according to the policy $\pi(\x,\u)$ that defines the probability of taking action $\u$ in 
state $\x$.

Considering the growth $\x_\mt$ as the state variable, the problem for the trader 
is to find the best strategy for buying and selling contracts $\ca$ and $\cb$, subject to 
the transaction costs $\zt_\mt$, in order to achieve the target short/medium term
growth $\xd$. The complete set of states for spoofing is determined for 
the state representation in \fref{f:gwSpoofing} as it captures the different levels of growth
after taking any of the actions. These actions are associated to the process of buying 
and selling contracts and are used by the trader to navigate in/within the state space of
\fref{f:gwSpoofing}, being the honest action set determined by
$\Uh=\{\uba,\usa,\usb,\ubb\}$ and similarly the set of 
manipulative actions $\Um=\{\usam,\ubam,\ubbm,\usbm\}$ 
(``Manipulative Buy $\ca$'', ``Manipulative Sell $\ca$'', ``Manipulative Sell $\cb$'',  
``Manipulative Buy $\cb$'', respectively), and the ``do nothing'' action for the ``buy 
and hold'' $\Un=\{\dn\}$, with $\U=\Uh\cup\Um\cup\Un$. The rewards are represented 
by the transaction costs $\zt_\mt$ that may depend on the action taken and the level of 
growth the trader is located at. The transition probabilities are linked to the degree of 
liquidity the contracts $\ca$ and $\cb$ have at a given tick $\mt$, so a good degree of 
liquidity will help the trader's orders to be filled and transition to a new level of growth, 
while low liquidity will restrict these transitions.

\subsection{\label{ss:pingingStrategy} Price Manipulation by Pinging}

Pinging is similar to spoofing as is defined as a limit order placed inside the
\textit{bid-ask spread} (the price difference between the best buy and sell quotes listed 
in the order book) without the intention of execution,  but cancelled almost instantly
\cite{scopino15}. This strategy is implemented by HFTs by exploiting the speed 
advantage with the intention to \textit{ping} the market in search of \textit{hidden 
liquidity}, \ie{orders that are not displayed in the order book as is the case of large 
orders placed by institutional investors}. This strategy has a more complex succession 
of events -- submit ping orders and almost instantly cancel them, detect hidden liquidity, 
take the liquidity on the trading side pursued by the large investor and then place the 
real orders at improved prices.

\begin{figure}[htp!]
\centering
\vspace{-5pt}
\includegraphics[scale=.8]{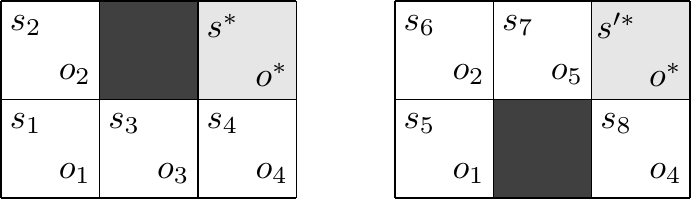}
\caption{Representation that illustrates pinging trading while trying to maximise the 
growth of the investment.}
\vspace{-10pt}
\label{f:gwPinging}
\end{figure}

In order to make pinging a successful strategy, HFTs must be able to find hidden
liquidity, a process that depends on the ping orders that help to create a belief on the 
existence of such liquidity. However, it is well known that investors prefer to place large 
orders in \textit{dark pools}, \ie{private venues where the exchange of assets is not 
visible to the general public, so no one can see who's buying/selling, but whose prices 
depends on the current market prices of well stablished markets}. This gives the
HFT uncertainty about the existence of hidden liquidity as it is not displayed in the 
order book. In order to simulate this in the representation of \fref{f:gwPinging},
we introduce the concept of observations that guides the trader on the actions to take 
while being on a given state. For example, having observation $\y_2$ while in level
$\x_2$ means that there is hidden liquidity (the obstacle) in the sell side of the $\cb$ 
contracts, so the HFT can produce profits by taking control over the prices in the 
regulated market while trading in the dark pool against the hidden liquidity. However, 
having the same observation in level $\x_6$ means that such liquidity does not exists 
and taking the manipulative action may produce losses.

\subsection{\label{ss:pomdpModel}Pinging as a Partially Observable Markov Decision 
Process}

Pinging, as described in \sref{ss:pingingStrategy}, can be modelled with a Partially 
Observable Markov Decision Process (\po\mdp)~\cite{baffaC2010}. In general,
the \po\mdp\ is defined by the tuple $\{\X,\U,\F,\J,\Y,\Om\}$, that is, the \mdp\ tuple is
extended with $\{\Y,\Om\}$, where $\Y$ represents a set of observations ($\y\in\Y$) and 
$\Om$ is a set of observation probabilities given states $\x$ and actions $\j$
($\{O(\y|\x,\j)\}$). For the \po\mdp, actions are taken according to the agent's belief of 
being on a given state and is calculated according to the observations.

Once more, every time a trader's order is filled then the correspondent transaction costs 
must be paid. Liquidity is again the one that facilitates the trading of contracts $\ca$ and 
$\cb$, so the trader can transition to the different levels of growth after the rebalance of 
capital. The observations represent the trader's detection of hidden liquidity (the 
obstacle) that may help or be counterproductive while seeking profits.

\section{\label{s:methodology}Methodology}

In both models, the trader has the objective to reach the goal $\x^*$ that represents 
the maximum investment growth and, since this is a bull market, the highest profit 
comes from having the most contracts, a process that can be performed by navigating 
within the state representations (the opposite also applies while in a \textit{bear market} 
[when pessimism persist and prices tend to fall], where the trader may prefer to sell 
contracts). We have chosen the state representations as in \fref{f:gwSpoofing} and 
\fref{f:gwPinging} as both model a single agent's behavior of acquiring contracts during 
a bull market period, with the option of taking either honest or manipulative actions as 
an ``optimal'' behavior. Other grids with a more complex structure may also reproduce 
optimality of trading strategies, but manipulative behavior may not emerge as an 
``optimal'' action according to the simulated market conditions, thus eliminating the core 
of the analysis we present in this paper.

Regardless of whether manipulative trading is permitted or not, the best sequence of 
trading actions for the agent (optimal policy) can be determined in a straightforward 
manner through, for example, reinforcement learning. In this paper, for the \mdp\ model 
this is achieved through simple value iteration \cite{suttonbarto1998} to find the optimal 
value function
\begin{equation}\label{e:valFunMDP}
\V^{\pi^*}(\x) = \max_{\u\in\U}\left[\J(\x,\u)+\gamma\sum_{\x'\in\X}P(\xp|\x,\u)\V(\x') 
\right],
\end{equation}
where $0<\gamma<$ is the discount factor. The \po\mdp\ formalism is intended to 
model states not fully observable, explaining why an observation function is 
needed to solve the problem. The observation function, $\Om(\u,\x,\y)$, is the 
probability of making observation $\y$ from state $\x$ after action
$\u$~\cite{kaelbling1998planning}. For \po\mdp's the solution is to find optimal policies 
with actions that maximises the value function. Based on the agent's current beliefs 
about the state (growth level), this value function can be represented as a system of 
simultaneous equations as
\begin{equation}\label{e:valFunPOMDP}
\V^{*}(\bs) = \max_{\u\in\U}\left[ \rho(\bs,\u)+\gamma\sum_{\bs'\in\Bss}\tau(\bs,\u,\bs')
\V^{*}(\bs') \right],
\end{equation}
where $\bs\in\Bss$ is a \textit{belief state}, $\rho(\bs,\u)=\sum_{\x\in\X}\bs(\x)\J(\x,\u)$ 
are the expected rewards for the belief states;
$\tau(\bs,\u,\bs')=\sum_{\{\y\in\Y|\bs=\bs'\}}P(\y|\u,\bs)$, the state transition function.

The optimal value function considers the potential rewards of actions taken in the 
future, so it captures the optimal actions that generate the most of rewards over the 
long-term. This argument enables us to examine the questions established at the end 
of \sref{s:intro} by analising the optimal actions in the (\po)\mdp\ model ultimately 
determined by two factors: the reward and the transition functions. Whether 
manipulative strategies are optimal, by expanding the idea of the reward function 
(transaction costs) and adding the notion of ``high fines/financial penalties'' imposed by 
market regulators, we argue this will encourage traders to stop the misbehavior and 
play in a fair way. Additionally, we believe that adding uncertainty to the effect of 
manipulation over liquidity may represent another cause to discourage abusive 
behavior and promote more efficient markets.

\section{\label{s:experiments}Experiments}

In this section we present a simulation on the profitability of the three strategies
described in \sref{s:probForm} and the optimal actions in the (\po)\mdp\ model in terms
of portfolio growth optimisation.

\subsection{\label{s:simulation}Simulation}

First, we simulate the ``buy and hold'', honest and spoofing strategies during bull 
market periods, by taking market indexes \textit{S\&P 500} and \textit{NASDAQ 
Composite} as the contracts $\ca$ and $\cb$, respectively. We use closing prices
for periods of 92 days, simulating a short term to produce profits.
\tref{t:simulationRes} shows the results of the simulation and we see that, in average,
spoofing outperforms the other strategies, a result that was produced by the increase of 
growth after taking manipulative actions. Honest trading is the second best strategy, 
beating the ``buy and hold'' as the later has good performance in the long term. The 
next step is to analyse at which point the optimality of manipulative actions appears in 
the (\po)\mdp\ models described in previous sections.
\begin{table}[t!]
\centering
\caption{Profitability simulated under a bull market.}
\scriptsize
\begin{tabular}{ |c|c|c| }
\hline
Strategy & Avg. profit (\%) & Std. dev. \\
\hline
Buy and Hold & 5.83\% & 0.041925 \\
Honest & 6.85\% & 0.059576 \\
Spoofing & 7.68\% & 0.059576\\
\hline
\end{tabular}
\vspace{-5pt}
\label{t:simulationRes}
\end{table}

\subsection{\label{ss:MDPsetup} \mdp\ Model for Spoofing}

\begin{enumerate}[leftmargin=0cm,itemindent=.6cm, label=\textit{\roman*.}]
\item\label{ss:spoofingOptimal}\textit{Is spoofing an optimal strategy?}

Here we demonstrate whether spoofing occurs according to the model described in 
\sref{ss:mdpModel} and, if it occurs, what are the factors that encourages it. First, we
try to model a market where all participants play under the same conditions -- all 
strategies pay the same transaction costs, with the trader's actions outcome totally
deterministic.

As a baseline, all honest actions in direction of the edge of any of the states have zero 
costs and make the agent bounce back to the same state. The same for manipulative 
actions, except that the obstacle switches its position (thus changing representation); 
otherwise, manipulative actions costs $-1$ in all states. Transitioning within the different 
states costs $-1$ and colliding against the obstacle has $0$/$-1$ costs for all honest/
manipulative actions. The terminal states, $\xd$ and $\x'^*$ have a reward of $+1$ 
meaning the trader has reached the desired growth state. The ``do nothing'' ($\dn$) 
action has zero costs in all states, but the agent cannot transition. All transitions are 
deterministic, meaning that $P(\xp|\x,\u)=1$, for all $\x,\u$.

We set $\gamma=0.95$ and solve equation \eref{e:valFunMDP} to find the optimal
actions in the \mdp\ model. \tref{t:polMDPmodel} shows the results for the baseline and
we see that spoofing do occur in most of the states, sharing optimality with honest 
actions. The results reveal that while trading under the same conditions in terms of 
transaction costs, spoofing can be exploited by traders in order to gain profits and reach 
the desired level of growth.

\item\label{ss:spoofFinPenalties}\textit{Adding financial penalties to spoofing}

Now, we try to encourage the trader to behave honestly by simulating market regulators 
imposing fines/financial penalties to spoofers. For this, in the baseline described in 
\sref{ss:MDPsetup}.\ref{ss:spoofingOptimal} we change the reward function and 
increase the costs to all manipulative actions in all states up to $-4.53$ and use the 
same value for $\gamma$ to solve \eref{e:valFunMDP}.

Once more, \tref{t:polMDPmodel} shows the results of this setup and the only optimal
actions are those associated to honest trading. There's a clear difference between the 
two market conditions from the regulatory point of view, one that considers a free-fine 
market (baseline) and the other with fines imposed to manipulators. For example, in the 
baseline starting from state $\x_2$ means for the spoofer takes only two steps to reach 
$\xd$ ($\usbm$ in $\x_2$; $\usbm$ in $\x_7$), while after imposing fines a honest 
trader will take four steps ($\usa$ in $\x_2$; $\ubb$ in $\x_1$; $\ubb$ in $\x_3$; and
$\uba$ in $\x_4$) to reach the same level of growth.

\item\label{ss:spoofUncerLiq}\textit{Adding uncertainty to liquidity}

A second attempt to stop the spoofing behavior is by providing uncertainty to the effects 
of manipulation over liquidity. This can be done by taking the baseline described in 
\sref{ss:MDPsetup}.\ref{ss:spoofingOptimal} and changing the transition function for all 
manipulative actions in all states. We take two different measures of uncertainty: a
50\%/50\%--10\%/90\% chance for the obstacle to switch/stagnates, with the aim to see 
which of these measures eliminates spoofing.

We take the same value for $\gamma$ and solve \eref{e:valFunMDP}. 
\tref{t:polMDPmodel} shows the results in this new setup and we conclude that 
implementing mechanism that somehow take control over liquidity are not as effective
as applying fines to manipulators. Spoofing still occurs despite the effects over liquidity 
are vague. However, we must notice that in both measures of uncertainty, almost all the 
optimal actions (including spoofing actions) are in the same direction of honest actions 
when adding fines to spoofers, meaning that the effect over liquidity is vague precisely 
because liquidity already exists in the market, a consistent result with our model 
described in \sref{ss:marketManipModel}.

\begin{table}[t!]
\centering
\caption{Optimal actions for the \mdp\ model under different conditions of the reward 
and transition functions.}
\scriptsize
\begin{tabular}{ |c|c|c|c|c| }
\hline
\multirow{2}{*}{State} & \multirow{2}{*}{Baseline} & Adding &
\multicolumn{2}{c|}{Adding uncertainty on liquidity} \\
 \cline{4-5}
 & & fines & 50\% vs. 50\%& 10\% vs. 90\% \\ 
 \hline
$\x_1$ & $\uba, \ubb, \usam$ &
      $\ubb$ &
      $\usbm$ &
      $\usbm$ \\
$\x_2$ & $\usbm$ &
      $\usa$ &
      $\usam, \ubbm$ &
      $\usa$ \\
$\x_3$ & $\ubb, \usam, \usbm$ &
      $\ubb$ &
      $\ubb, \usbm$ & 
      $\ubb, \usbm$ \\
$\x_4$ & $\uba, \usam$ &
      $\uba$ &
      $\uba, \usam$ &
      $\uba, \usam$ \\
$\x_5$ & $\uba, \usam, \usbm$ &
      $\uba$ &
      $\uba$ &
      $\uba$ \\
$\x_6$ & $\ubb$ &
      $\ubb$ &
      $\ubb$ &
      $\usbm$ \\
$\x_7$ & $\ubb, \usbm$ &
      $\ubb$ &
      $\ubb,\usbm$ &
      $\ubb, \usbm$ \\
$\x_8$ & $\uba, \usam$ &
      $\uba$ &
      $\uba, \usam$ &
      $\uba, \usam$ \\
\hline
\end{tabular}
\vspace{-5pt}
\label{t:polMDPmodel}
\end{table}

\end{enumerate}

\subsection{\label{ss:POMDPsetup} \po\mdp\ Model for Pinging}

\begin{enumerate}[leftmargin=0cm,itemindent=.6cm, label=\textit{\roman*.}]
\item\label{ss:pingingOptimal}\textit{Is pinging an optimal strategy?}

Here we demonstrate whether pinging emerges as an optimal strategy while maximising 
growth. First, we use the same baseline as in
\sref{ss:MDPsetup}.\ref{ss:spoofingOptimal} in terms of rewards and transitions and the
observations shown in \fref{f:gwPinging}. We set $\gamma=0.95$ and solve 
\eref{e:valFunPOMDP}. \tref{t:polPOMDPmodel} shows the results for the baseline in 
the \po\mdp\ model, where pinging is the optimal action in all observable states under 
equal conditions in terms of transaction costs. Honest behavior is optimal only in 
observed state $\y_2$, meaning that no matter the HFT actually observes (detects) the 
hidden liquidity, she will take profits from other investors that do not necessarily place 
large orders.

\item\label{ss:pingTransCosts}\textit{Increasing transaction costs to pinging}

We want to discourage HFTs to take pinging by changing parameters that may 
influence trader's decisions. As pinging is not considered illegal, we take the changes in 
the reward function equivalent to increasing the direct transaction costs associated to 
pinging. We change the reward function for the \po\mdp\ model in the baseline 
considered in \sref{ss:POMDPsetup}.\ref{ss:pingingOptimal} and increase the costs to 
all pinging actions up to $-4.91$ in all states. We set $\gamma=0.95$ and solve 
\eref{e:valFunPOMDP}.

\tref{t:polPOMDPmodel} shows the results for these changes and we see that, under 
the new conditions pinging is no longer optimal as is was in the baseline in
\sref{ss:POMDPsetup}.\ref{ss:pingingOptimal}, and only honest actions can be taken by
the trader. This means that the core of the business related to pinging is no longer 
profitable because of the high costs that must be paid to the correspondent parties -- it
may be the case that pinging produce profits, but not large enough to cover the 
transaction costs.

\item\label{ss:pingUncerLiq}\textit{Adding uncertainty to liquidity}

Finally, a second attempt to stop pinging trading is by changing the transition function,
a mechanism that applies uncertainty to the potential effects of pinging over market
liquidity. We take the baseline as in \sref{ss:POMDPsetup}.\ref{ss:pingingOptimal} 
and the same transitions described in \sref{ss:MDPsetup}.\ref{ss:spoofUncerLiq}. We 
set $\gamma=0.95$ and solve \eref{e:valFunPOMDP}. Once more, 
\tref{t:polPOMDPmodel} shows the results of these changes and we see that, under
mechanism that provide uncertainty to the effect of pinging trades over liquidity, pinging
is still an optimal action in some of the observed states, showing that is more effective
to increase the transaction costs as shown in
\sref{ss:POMDPsetup}.\ref{ss:pingTransCosts}, a similar result as in spoofing.

\begin{table}[t!]
\centering
\caption{Optimal actions for the \po\mdp\ model under different conditions of the reward 
and transition functions.}
\scriptsize
\begin{tabular}{ |c|c|c|c|c| }
\hline
Observed & \multirow{2}{*}{Baseline} & Increase &
\multicolumn{2}{c|}{Adding uncertainty on liquidity} \\
 \cline{4-5}
State & & transaction costs & 50\% vs. 50\%& 10\% vs. 90\% \\ 
 \hline
$\y_1$ & $\usam$ &
      $\ubb$, $\uba$ &
      $\uba$ &
      $\uba$, $\usbm$ \\
$\y_2$ & $\usbm$, $\ubb$ &
      $\ubb$, $\usa$ &
      $\ubb$, $\usbm$ &
      $\ubb$, $\usbm$, $\usa$ \\
$\y_3$ & $\usam$ &
      $\ubb$ &
      $\ubb$ &
      $\ubb$ \\
$\y_4$ & $\usam$ &
      $\uba$ &
      $\uba$ &
      $\uba$ \\
$\y_5$ & $\usbm$ &
      $\ubb$ &
      $\ubb$ &
      $\ubb$ \\
\hline
\end{tabular}
\vspace{-5pt}
\label{t:polPOMDPmodel}
\end{table}

\end{enumerate}

\section{\label{sec:conclusions}Conclusions}

The results from the (\po)\mdp\ models show they can predict behaviours, and both the 
manipulative and honest trading can co-exist in a regulated market where all
participants have the same direct costs. We found that both spoofing and pinging 
trading are optimal investment strategies while traders try to maximise the investment 
growth, but market regulators can discourage the use of these strategies by 
implementing mechanism over market liquidity, and this enforcement will be more 
efficient if fines are added (for spoofing) or by increasing the direct transaction costs 
(pinging).

However, our model works on bull market conditions and we expect to fit on bear 
markets if we change the side of the trading actions. Other conditions where no trends 
exists may produce incentives for manipulation as a way to move the market.
Furthermore, in pinging HFTs have the option to avoid ping orders  and analyse the 
predictability of the asset's order flow with the goal to infer the existence of hidden 
liquidity, thus saving direct transaction costs.

Further research can be focused on applying the models in real market data and more
complex portfolios, and verify the effectiveness of the recommendations 
provided to disincent manipulation performed by spoofing/pinging traders.

%

\newpage
\bibliography{paper.bib}
\bibliographystyle{aaai}

\end{document}